\documentclass[aps,superscriptaddress,superbib,twocolumn,groupedaddress]{revtex4}
\usepackage[final]{neel}

\bibpunct{[}{]}{,}{n}{}{}

\usepackage[latin1]{inputenc}
\usepackage[T1]{fontenc}

\def\figurename{Fig.}

\begin{document}

\renewcommand\topfraction{0.8}
\renewcommand\bottomfraction{0.7}
\renewcommand\floatpagefraction{0.7}

\title{Observation of Bloch-point domain walls in cylindrical magnetic nanowires}%

\author{S. Da Col}
\affiliation{Univ. Grenoble Alpes, Inst NEEL, F-38042 Grenoble, France}%
\affiliation{CNRS, Inst NEEL, F-38042 Grenoble, France}

\author{S. Jamet}
\affiliation{Univ. Grenoble Alpes, Inst NEEL, F-38042 Grenoble, France}%
\affiliation{CNRS, Inst NEEL, F-38042 Grenoble, France}

\author{N. Rougemaille}
\affiliation{Univ. Grenoble Alpes, Inst NEEL, F-38042 Grenoble, France}%
\affiliation{CNRS, Inst NEEL, F-38042 Grenoble, France}

\author{A. Locatelli}
\affiliation{Elettra - Sincrotrone Trieste S.C.p.A., I-34012 Basovizza, Trieste, Italy}%

\author{T. O. Mentes}
\affiliation{Elettra - Sincrotrone Trieste S.C.p.A., I-34012 Basovizza, Trieste, Italy}%

\author{B. Santos Burgos}
\affiliation{Elettra - Sincrotrone Trieste S.C.p.A., I-34012 Basovizza, Trieste, Italy}%

\author{R. Afid}
\affiliation{Univ. Grenoble Alpes, Inst NEEL, F-38042 Grenoble, France}%
\affiliation{CNRS, Inst NEEL, F-38042 Grenoble, France}

\author{M. Darques}%
\affiliation{Univ. Grenoble Alpes, Inst NEEL, F-38042 Grenoble, France}%
\affiliation{CNRS, Inst NEEL, F-38042 Grenoble, France}
\altaffiliation{Now at: Nexans research center, 69007 Lyon, France}

\author{L. Cagnon}
\affiliation{Univ. Grenoble Alpes, Inst NEEL, F-38042 Grenoble, France}%
\affiliation{CNRS, Inst NEEL, F-38042 Grenoble, France}

\author{J. C. Toussaint}
\affiliation{Univ. Grenoble Alpes, Inst NEEL, F-38042 Grenoble, France}%
\affiliation{CNRS, Inst NEEL, F-38042 Grenoble, France}

\author{O. Fruchart}
\affiliation{Univ. Grenoble Alpes, Inst NEEL, F-38042 Grenoble, France}%
\affiliation{CNRS, Inst NEEL, F-38042 Grenoble, France}
\email[]{Olivier.Fruchart@grenoble.cnrs.fr}

\date{\today}


\begin{abstract}

Topological protection is an elegant way of warranting the integrity of quantum and nanosized systems. In magnetism one example is  the Bloch-point, a peculiar object implying the local vanishing of magnetization within a ferromagnet. Its existence had been postulated and described theoretically since several decades, however it has never been observed. We confirm experimentally the existence of Bloch points, imaged within domain walls in cylindrical magnetic nanowires, combining surface and transmission XMCD-PEEM magnetic microscopy. This opens the way to the experimental search for peculiar phenomena predicted during the motion of Bloch-point-based domain walls.

\end{abstract}

\maketitle


There is a rising interest for physical systems providing topological protection. The interest is both fundamental to elucidate the underlying physical phenomena, and applied as a mean to provide robustness to a state against external perturbations and decoherence pathways. For example, the peculiar topology of the band structure of carbon nanotubes and graphene forbids backscattering of charge carriers\cite{bib-AND1998}, an effect which is often invoked to explain the high mobilities up to room temperature\cite{bib-NOV2007}. A similar effect occurs at the surface of so-called topological insulators, together with a locking of the spin of charge carriers; this provides spin currents protected against depolarization\cite{bib-KOeN2007}. A photonic analogue was also designed by combining helical wave guides on a lattice with a graphene-like honeycomb topology, removing time-reversal symmetry and thereby preventing backscattering of light\cite{bib-REC2013}.

In systems displaying a directional order parameter such as liquid crystals and ferromagnets, interesting phenomena are associated with the slowly-varying texture of the order field (magnetization for a ferromagnet). The requirement of local continuity of a vector field with fixed magnitude provides a topological protection against the transformation of the texture. A prototypical case in magnetism is skyrmions, which are essentially local two-dimensional chiral spin textures stabilized by the Dzyaloshinskii-Moriya interaction, embedded in an otherwise uniformly-magnetized surrounding. Despite these surroundings skyrmions cannot unwind continuously as explained by the above continuity constraints of the magnetization field, explaining their topological protection. Skyrmions have first been predicted theoretically\cite{bib-BOG1989}, then confirmed experimentally in both bulk\cite{bib-YU2010} and thin film forms\cite{bib-ROM2013}.

\begin{figure}
  \begin{center}
  \includegraphics[width=87.250mm]{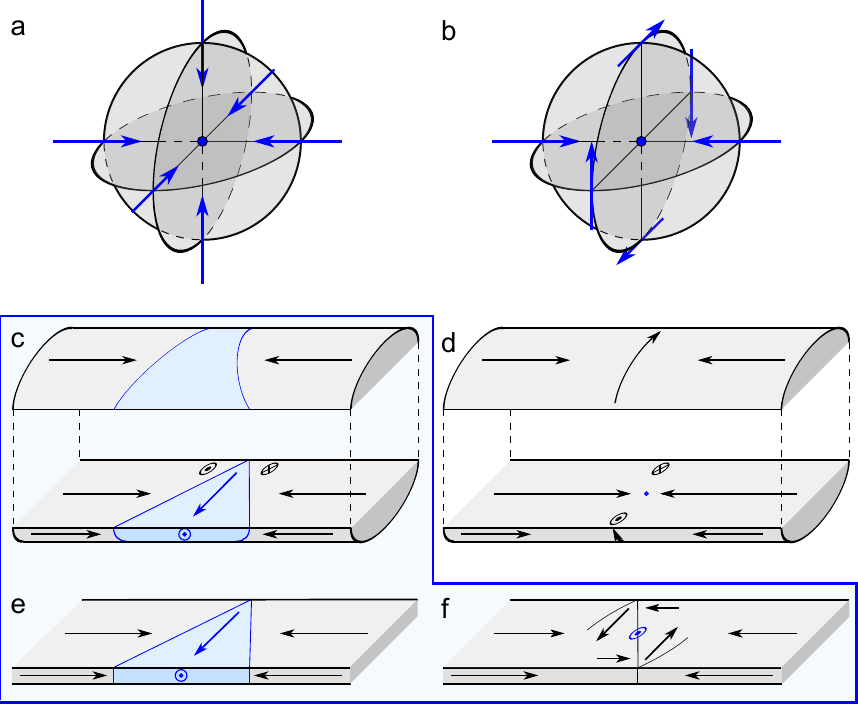}%
  \caption{\label{fig-dw-schematics}Sketches for magnetization textures. (a-b)~Two of the various possible types of Bloch points. (b) is obtained from (a) with a quarter turn of magnetization around the $x$ axis (c-d)~Cylindrical wires with DWs of type transverse and Bloch-point, respectively. (e-f)~Strips with in-plane magnetization with DWs of type transverse and vortex, respectively. Notice the identical boundary conditions around the Bloch Point in b and d. DWs (c,d,f) share the same topology, characterized by an area of induction going through the system along a transverse direction, depicted in blue. The Bloch-point DW is of a different class, and cannot be mapped continuously to the wire TW.}
  \end{center}
\end{figure}

Bloch points are yet another type of topologically-protected magnetic texture which cannot be unwound, however of a three-dimensional nature. Bloch points are such that given the distribution of magnetization set on a closed surface like a sphere, the enclosed volume cannot be mapped with a continuous magnetization field of finite magnitude. This occurs \eg for hedge-hog configurations, or more generally whenever all directions of magnetization are mapped on the closed surface\bracketsubfigref{fig-dw-schematics}{a-b}. Such boundary conditions imply the local cancellation of the modulus of magnetization on at least one location, which is a singularity for a ferromagnetic material. Although of nearly atomic size because of the cost in exchange energy, the Bloch point is topologically protected by its extended boundary conditions. Bloch points were predicted theoretically\cite{bib-FEL1965,bib-DOE1968}, their existence being suspected from the examination of boundary conditions at the surfaces of three-dimensional samples such as former bubble-memory media\cite{bib-MAL1979}.

The interest in Bloch points was revived in recent years. As a zero-dimensional object, they were predicted to be required in the transient state allowing magnetization reversal along one-dimensional objects, such as magnetic vortices\cite{bib-THI2003}. Due to their atomic size Bloch points are thought to be subject to pinned at atomic sites on the underlying crystal\cite{bib-THI2003,bib-KIM2013,bib-PIA2013}. Based on the interaction with these pinning sites, it was postulated that moving Bloch points emit THz waves. Bloch points have also been predicted to exist at rest in magnetic nanowires with a compact cross-section such as cylindrical. In nanowires two types of DWs have indeed been predicted to exist: the transverse wall (TW) and the Bloch-Point wall(BPW)\cite{bib-FOR2002,bib-THI2006}\bracketsubfigref{fig-dw-schematics}{c-d}. The former is topologically equivalent to the transverse and vortex walls in flat strips\bracketsubfigref{fig-dw-schematics}{e-f}, in which a flux of induction runs across the structure. To the contrary, the latter has a distinct topology: magnetization remains mostly locally parallel the the surface at any location, allowing to decrease the magnetostatic energy. In other words, no flux line flows across the wire. As described above, these boundary conditions impose the existence of a Bloch point, which happens to sit on the axis of the wire. The wall of lowest energy should be the TW for wire diameter smaller than roughly seven times the dipolar exchange length $\DipolarExchangeLength=\sqrt{2A/\muZero\Ms^2}$, while the BPW should be the ground state for larger diameters. Nevertheless, each type of DW should exist as a metastable state over a significant range of diameters.

The existence of BPWs would have practical macroscopic consequences: micromagnetic simulation predicts that the topological protection of the BPW prevents its transformation into other types of domain walls (DWs) during its motion\cite{bib-FOR2002,bib-THI2006}. As a consequence domain-wall speeds beyond $\unit[1]{\kilo\meter\per\second}$ should be reachable, opening the way to new physics such as the spin-Cherenkov effect through interaction of the domain wall with standing spin-waves\cite{bib-YAN2011b}. This is in strong contrast with the more common case of flat strips made by lithography, for which the two possible DWs share a common topology, so that periodic transformation from one to another during domain-wall motion severely limits the average mobility, and motion is non-stationary\cite{bib-THI2006,bib-THI2008}.

The existence of BPWs has not been confirmed experimentally yet. In this Letter we use a three-dimensional high-spatial resolution magnetic imaging technique to gather both surface and volume information of the magnetization texture of domain walls in cylindrical nanowires, and formally identify the BPW. These results are supported by the development of a post-processing code of three-dimensional magnetization textures to analyze the experimental magnetic contrast.

We first prepared self-organized anodized alumina templates in $\unit[0.3]{M}$ oxalic acid at either $\unit[40]{V}$ or $\unit[135]{V}$, yielding pore diameters $\unit[35]{\nano\meter}$ and $\unit[120]{\nano\meter}$, respectively\cite{bib-LEE2006}. For some templates we modulated the voltage from $\unit[135]{V}$ to $\unit[150]{V}$ so as to vary the diameter of pores along their length\cite{bib-LEE2008}. We also made use of atomic layer deposition to reduce uniformly the diameter along their length. We then electroplated $\mathrm{Fe}_{20}\mathrm{Ni}_{80}$ (Permalloy) micrometers-long nanowires in these. The filled alumina templates were dissolved in a NaOH solution, followed by several rinsing steps in water and finally in isopropyl-alcohol. Drops of solution were deposited on doped Si wafers. To identify the nature of the DWs we applied element-sensitive X-ray circular magnetic dichroism photoemission electron microscopy measurements (XMCD-PEEM), carried out using the setup\cite{bib-LOC2006} operating at the undulator beamline Nanospectroscopy at Elettra, Sincrotrone Trieste. The photons impinge on the supporting surface with a grazing angle of $\angledeg{16}$. In this Letter, we present data based on secondary photoelectron emission at the Fe $\mathrm{L}_{3}$ edge using mostly circularly polarized radiation as a probe. Magnetic contrast was obtained by difference of images with opposite helicities $\sigma_{+}$ and $\sigma_{-}$ of the photon beam, normalized to their sum XPEEM. The XMCD image intensity is thus given by $I_{\mathrm{XMCD}} = (I_{\sigma_{-}} - I_{\sigma_{+}}) / (I_{\sigma_{-}} + I_{\sigma_{+}}$), with the convention of positive contrast for magnetization parallel to the incoming beam. The spatial resolution in XMCD-PEEM mode is $\approx\unit[30]{\nano\meter}$. Micromagnetic simulations were performed using feellgood, a home-built code based on the temporal integration of the Landau-Lifshitz-Gilbert equation in a finite element scheme, \ie using tetrahedra to discretize the nanowires\cite{bib-ALO2012}. Only exchange and magnetostatic interactions were taken into account, to deal with the present case of magnetically-soft wires. The parameters for bulk permalloy were used: $A=\unit[10^{-11}]{\joule\per\meter}$ and $\muZero\Ms=\unit[1]{\tesla}$. The tetrahedron size was about \unit[4]{\nano\meter}.

Whereas magnetic nanowires gave rise to a numerous literature\cite{bib-FER1999a}, essentially macroscopic and full magnetization reversal was probed\cite{bib-WAN2009}. One reason is that magnetization is constrained to remain essentially along the wire due to magnetostatics. Magnetization switching thus proceeds under a significant applied magnetic field through nucleation of a domain wall at one end, followed by its fast motion towards its other end, leading to its annihilation. To stabilize domain walls in nanowires we make use of local protrusions placed every other several micrometers, to act as potential barriers for domain walls motion. While these barriers are not sufficient to pin domain walls when the magnetic field is applied along the wire\cite{bib-PIT2011}, we found that they are fit to prevent motion of domain walls upon dc oscillatory demagnetization with the magnetic field applied perpendicular to the wafer plane.

\begin{figure}
  \begin{center}
  \includegraphics[width=83.283mm]{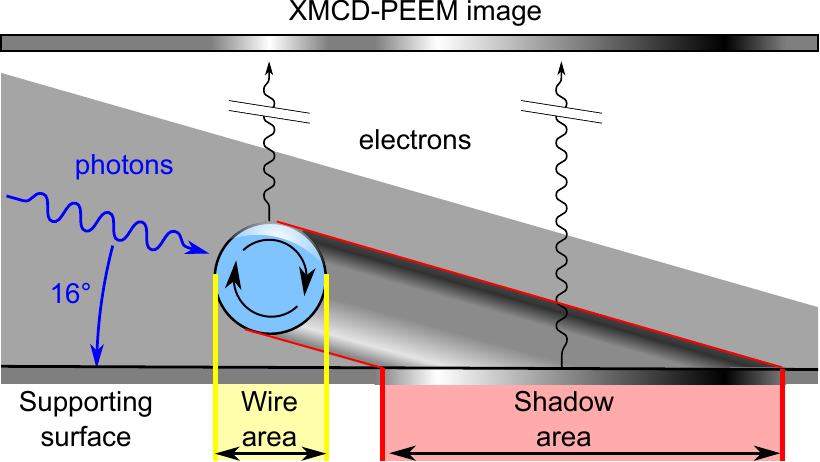}%
  \caption{\label{fig-dw-absorption}Schematic of the principle of the dual surface transmission-projection PEEM, with magnetic sensitivity based on XMCD. The grey level stands for the dichroic signal (normalized difference with photons of opposite helicity).}
  \end{center}
\end{figure}

The type of domain walls thus created was examined by XMCD-PEEM. Thanks to the three-dimensional nature of nanowires we could collect both direct photo-emission as usually done, and also transmission data as shown on \figref{fig-dw-absorption}. Direct photoemission provides magnetic contrast on the wires. As the secondary electrons collected by the microscope have a mean free path of a few nanometers only, this contrast informs us on surface magnetization only. We also analyzed the magnetic contrast formed in the shadow by the light partially transmitted through the wire\cite{bib-KIM2011b}. This allows us to gain information about the magnetization arrangement in the bulk of the wires, integrated along the path of light. Gathering this information is crucial as identification of the DW type may otherwise be ambiguous if relying solely from the map of surface magnetization. Besides, this allows us to check the arrangement of magnetization on the wire axis, where Bloch points are expected. Finally, notice that thanks to the rather grazing incidence $\alpha=\unit[16]{\degree}$ the shadow is inflated with respect to the wire diameter by a factor $1/\sin\alpha\approx3.6$, thereby potentially bringing the spatial resolution of the microscope in this projection mode around $\unit[10]{\nano\meter}$.

\begin{figure}
  \begin{center}
  \includegraphics[width=83.227mm]{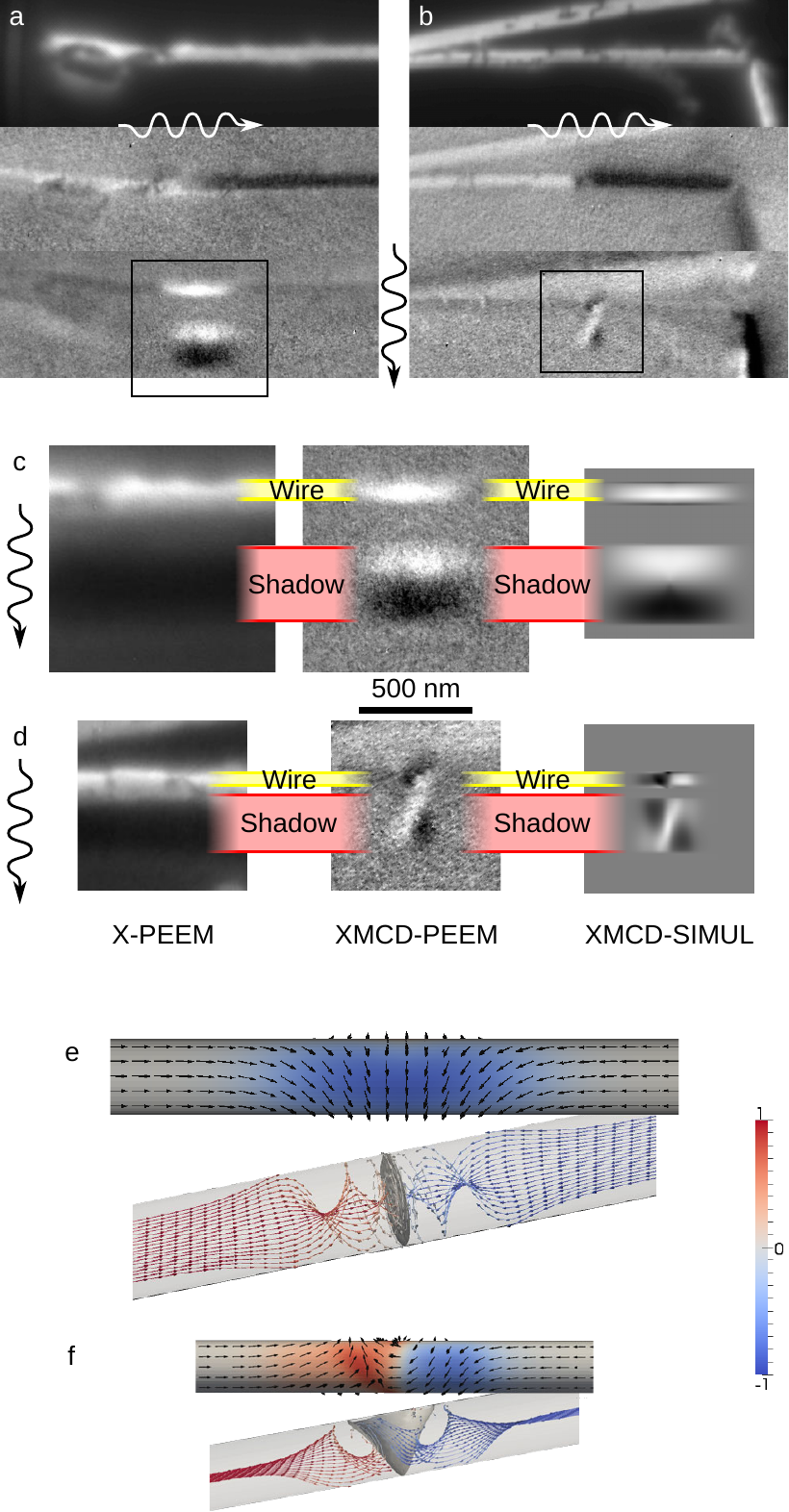}%
  \caption{\label{fig-dw-identification}Identification of the Bloch-point and transverse domain walls based on XMCD-PEEM compared with simulations.\dataref{TW: Eowyn and Ezechiel; BPW: Edgar, second DW (MD2013-01-19-042)} (a-b)~In each column are three views of the same wire, at the same location and with a field of view $\unit[3\times1]{\micro\meter}$. From top to bottom total absorbtion, and XMCD contrast parallel and across the wires. (c-d)~From left to right: X-PEEM (sum of images for the two polarizations), XMCD-PEEM (enlargement of the square areas in (a-b) and simulations of two DWs at the Fe L3 edge. The photons arrive from the upper part of the images. (c)~A wire of diameter $\unit[95]{\nano\meter}$ lifted $\unit[80]{\nano\meter}$ above the surface, with a DW identified as of Bloch-point type. (d)~A wire of diameter $\unit[70]{\nano\meter}$ lifted $\unit[25]{\nano\meter}$ above the surface, with a DW identified as of transverse type. (e-f)~top-view and open view of the micromagnetic state used in the right parts of c-d.}
  \end{center}
\end{figure}

In a first step the wires were aligned along the photon beam so as to identify longitudinal domains, and thus highlight the location of domain walls. In a second step the sample is rotated by $\unit[90]{\degree}$ so that the wires are aligned in the direction transverse to the photon beam. Under this configuration contrast solely arises from domain walls\bracketsubfigref{fig-dw-identification}{a-b}. We observed two well-defined families of DWs, typical examples being shown in \figref{fig-dw-identification}{c-d}. The first family is characterized by an orthoradial curling of magnetization as identified from the shadow, and it is symmetric with respect to a plane perpendicular to the wire axis. Notice the absence of contrast on the axis, as expected for the presence of a BP. The second type of DW breaks the above-mentioned symmetry, and is now characterized by a monopolar contrast at the center of the DW on the axis of the wire. This is expected for TWs, the contrast on the axis arising from the transverse component of magnetization with respect to the beam. Consistently with predictions\cite{bib-THI2006}, the family ascribed to BPWs were found in wires with larger diameter than for TWs.

We developed simulations to allow for a quantitative analysis of the experimental contrasts, and strengthen the symmetry arguments provided above. Starting from crude micromagnetic distributions similar those two depicted in \subfigref{fig-dw-schematics}{c-d}, the system is let to evolve and dissipate energy to finally reach a local minimum in either the TW or BPW state\bracketfigref{fig-dw-identification}{e-f}. These configurations are then post-processed to deliver a simulation of XMCD-PEEM contrast. To do this the wire is intercepted with a dense beam of parallel lines, each acting as the trace of a photon. First, for each photon helicity the probabilistic absorption of any trace is integrated along its path through the nanowire. The absorption coefficient per unit length is calculated depending on the local direction of magnetization with respect to the propagation vector, and is different for the two photon helicities. Second, the photoemission process is simulated. Given the small mean free path of the low-energy electrons collected in the microscope, we considered only those electrons emitted at the very surface, proportionally to the local absorbtion. At the surface of the magnetic wire we computed $I_{\sigma_{-}}$ and $I_{\sigma_{+}}$ as the intensity of each beam helicity multiplied by the magnetization-dependent absorption, for both incoming and outgoing photons. However, at the surface of the non-magnetic supporting surface where the shadow is projected, the difference signal only reflects the imbalance of the photon intensities.

\begin{figure}
  \begin{center}
  \includegraphics[width=61.131mm]{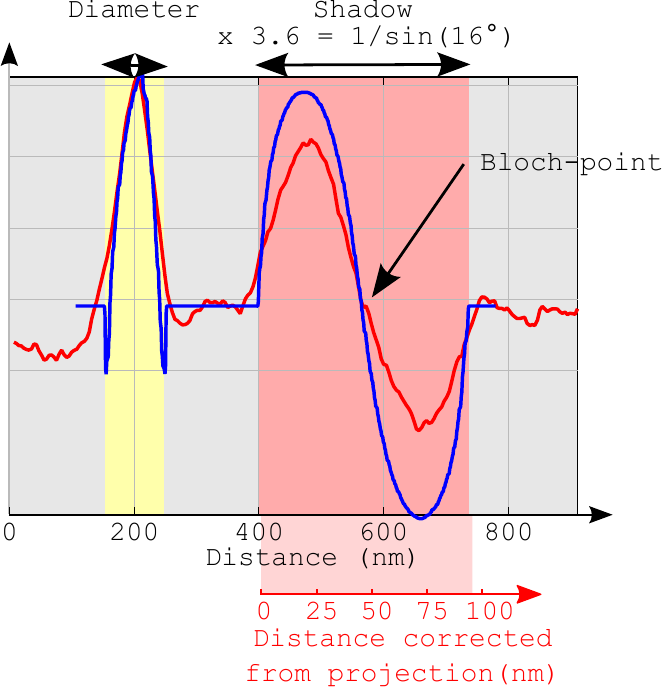}%
  \caption{\label{fig-dw-quantitative}Comparison of the experimental and simulated cross-section of the contrast over a Bloch-point domain wall. Data from \figref{fig-dw-identification}}
  \end{center}
\end{figure}

The simulations reproduce all features of the experimental contrast, thereby confirming the observation of both TW and BPW in nanowires. While TWs were imaged independently recently\cite{bib-BIZ2013}, this is the first experimental confirmation of the existence of BPWs. The agreement of simulations with experiments is quantitative, see for example the rapid increase of width of DW with the diameter of the nanowire\bracketsubfigref{fig-dw-identification}{c}, or the exact shape across the BPW\bracketfigref{fig-dw-quantitative} with zero contrast at the expected location of the Bloch point. As regards the TW, the black and white features at the surface of the wire is shown to result from the curling of magnetization around the transverse core.

The experimental demonstration of the existence of BPWs opens the way for the investigation of its peculiar behavior of motion, predicted numerically. We already mentioned its expected steady motion and high velocity, limited only by the crossing of the spin-wave dispersion curve\cite{bib-YAN2011b}. This regime has never been evidenced in any type of domain wall, as domain walls investigated so far undergo periodic transformations or have a too low mobility to reach this regime. The interaction of the Bloch point with the underlying lattice may also provide a nano-source of THz, with frequency tuneable through the DW speed, itself controlled by the applied magnetic field. Other features have been predicted, such as the once-only switch of the chirality of its orthoradial curling, selected depending on the two parameters: charge of the domain wall (head-to-head or tail-to-tail) and direction of motion\cite{bib-THI2006}. Tackling these phenomena requires that domain walls are not too much pinned on local defects. We have investigated the response of both types of DWs to quasistatic pulses of magnetic field, inspecting the DWs with magnetic force microscopy. The propagation field remains moderate, in the range $\unit[1\mathrm{-}10]{\milli\tesla}$. This is comparable to \eg that of perpendicularly-magnetized strips, where DW motion could be extensively investigated\cite{bib-MOO2008,bib-MIR2011}. These wires should therefore be already suitable to confirm the predictions for motion of BPWs\cite{bib-THI2006,bib-YAN2010}, while progress in material control are expected after this first demonstration. Besides fundamental studies, this should also redraw attention to the proposal of a three-dimensional race-track memory based on 2D arrays of parallel cylindrical nanowires\cite{bib-PAR2004,bib-PAR2008}.



\section*{Acknowledgements}

The research leading to these results has received funding from the European Unions's 7th Framework Programme under grant agreement n°309589 (M3d). We gratefully acknowledge the help of E.~Wagner for the demagnetization of samples, Ph.~David for the preparation of MFM tips, Simon Le Denmat for technical support with AFM/MFM, and fruitful discussions with  J.~Vogel, S.~Pizzini, J.~Coraux and M.~Sta\u{n}o.

\section*{References}


\begin{thebibliography}{0}
\expandafter\ifx\csname natexlab\endcsname\relax\def\natexlab#1{#1}\fi
\expandafter\ifx\csname bibnamefont\endcsname\relax
  \def\bibnamefont#1{#1}\fi
\expandafter\ifx\csname bibfnamefont\endcsname\relax
  \def\bibfnamefont#1{#1}\fi
\expandafter\ifx\csname citenamefont\endcsname\relax
  \def\citenamefont#1{#1}\fi
\expandafter\ifx\csname url\endcsname\relax
  \def\url#1{\texttt{#1}}\fi
\expandafter\ifx\csname urlprefix\endcsname\relax\def\urlprefix{URL }\fi
\providecommand{\bibinfo}[2]{#2}
\providecommand{\eprint}[2][]{\url{#2}}

\end{thebibliography}


\begin{thebibliography}{31}
\expandafter\ifx\csname natexlab\endcsname\relax\def\natexlab#1{#1}\fi
\expandafter\ifx\csname bibnamefont\endcsname\relax
  \def\bibnamefont#1{#1}\fi
\expandafter\ifx\csname bibfnamefont\endcsname\relax
  \def\bibfnamefont#1{#1}\fi
\expandafter\ifx\csname citenamefont\endcsname\relax
  \def\citenamefont#1{#1}\fi
\expandafter\ifx\csname url\endcsname\relax
  \def\url#1{\texttt{#1}}\fi
\expandafter\ifx\csname urlprefix\endcsname\relax\def\urlprefix{URL }\fi
\providecommand{\bibinfo}[2]{#2}
\providecommand{\eprint}[2][]{\url{#2}}

\bibitem[{\citenamefont{Ando et~al.}(1998)\citenamefont{Ando, Nakanishi, and
  Saito}}]{bib-AND1998}
\bibinfo{author}{\bibfnamefont{T.}~\bibnamefont{Ando}},
  \bibinfo{author}{\bibfnamefont{T.}~\bibnamefont{Nakanishi}},
  \bibnamefont{and} \bibinfo{author}{\bibfnamefont{R.}~\bibnamefont{Saito}},
  \bibinfo{journal}{\jpsj} \textbf{\bibinfo{volume}{67}}, \bibinfo{pages}{2857}
  (\bibinfo{year}{1998}).

\bibitem[{\citenamefont{Novoselov et~al.}(2007)\citenamefont{Novoselov, Zhang,
  Morozov, Stormer, Zeitler, Maan, Boebinger, Kim, and Geim}}]{bib-NOV2007}
\bibinfo{author}{\bibfnamefont{K.~S.} \bibnamefont{Novoselov}},
  \bibinfo{author}{\bibfnamefont{Z.~J.~Y.} \bibnamefont{Zhang}},
  \bibinfo{author}{\bibfnamefont{S.~V.} \bibnamefont{Morozov}},
  \bibinfo{author}{\bibfnamefont{H.~L.} \bibnamefont{Stormer}},
  \bibinfo{author}{\bibfnamefont{U.}~\bibnamefont{Zeitler}},
  \bibinfo{author}{\bibfnamefont{J.~C.} \bibnamefont{Maan}},
  \bibinfo{author}{\bibfnamefont{G.~S.} \bibnamefont{Boebinger}},
  \bibinfo{author}{\bibfnamefont{P.}~\bibnamefont{Kim}}, \bibnamefont{and}
  \bibinfo{author}{\bibfnamefont{A.~K.} \bibnamefont{Geim}},
  \bibinfo{journal}{\science} \textbf{\bibinfo{volume}{315}},
  \bibinfo{pages}{1379} (\bibinfo{year}{2007}).

\bibitem[{\citenamefont{K\"{o}nig et~al.}(2007)\citenamefont{K\"{o}nig, Wiedmann,
  Br\"{u}ne, Roth, Buhmann, Molenkamp, Qi, and Zhang}}]{bib-KOeN2007}
\bibinfo{author}{\bibfnamefont{M.}~\bibnamefont{K\"{o}nig}},
  \bibinfo{author}{\bibfnamefont{S.}~\bibnamefont{Wiedmann}},
  \bibinfo{author}{\bibfnamefont{C.}~\bibnamefont{Br\"{u}ne}},
  \bibinfo{author}{\bibfnamefont{A.}~\bibnamefont{Roth}},
  \bibinfo{author}{\bibfnamefont{H.}~\bibnamefont{Buhmann}},
  \bibinfo{author}{\bibfnamefont{L.~W.} \bibnamefont{Molenkamp}},
  \bibinfo{author}{\bibfnamefont{X.-L.} \bibnamefont{Qi}}, \bibnamefont{and}
  \bibinfo{author}{\bibfnamefont{S.-C.} \bibnamefont{Zhang}},
  \bibinfo{journal}{\science} \textbf{\bibinfo{volume}{318}},
  \bibinfo{pages}{766} (\bibinfo{year}{2007}).

\bibitem[{\citenamefont{Rechtsman et~al.}(2013)\citenamefont{Rechtsman, Zeuner,
  Plotnik, Lumer, Podolsky, Dreisow, Nolte, Segev, and Szameit}}]{bib-REC2013}
\bibinfo{author}{\bibfnamefont{M.~C.} \bibnamefont{Rechtsman}},
  \bibinfo{author}{\bibfnamefont{J.~M.} \bibnamefont{Zeuner}},
  \bibinfo{author}{\bibfnamefont{Y.}~\bibnamefont{Plotnik}},
  \bibinfo{author}{\bibfnamefont{Y.}~\bibnamefont{Lumer}},
  \bibinfo{author}{\bibfnamefont{D.}~\bibnamefont{Podolsky}},
  \bibinfo{author}{\bibfnamefont{F.}~\bibnamefont{Dreisow}},
  \bibinfo{author}{\bibfnamefont{S.}~\bibnamefont{Nolte}},
  \bibinfo{author}{\bibfnamefont{M.}~\bibnamefont{Segev}}, \bibnamefont{and}
  \bibinfo{author}{\bibfnamefont{A.}~\bibnamefont{Szameit}},
  \bibinfo{journal}{\nature} \textbf{\bibinfo{volume}{496}},
  \bibinfo{pages}{196} (\bibinfo{year}{2013}).

\bibitem[{\citenamefont{Bogdanov and Yablonskii}(1989)}]{bib-BOG1989}
\bibinfo{author}{\bibfnamefont{A.~N.} \bibnamefont{Bogdanov}} \bibnamefont{and}
  \bibinfo{author}{\bibfnamefont{D.~A.} \bibnamefont{Yablonskii}},
  \bibinfo{journal}{\spJETP} \textbf{\bibinfo{volume}{68}},
  \bibinfo{pages}{101} (\bibinfo{year}{1989}).

\bibitem[{\citenamefont{Yu et~al.}(2010)\citenamefont{Yu, Onose, an~J.~H.~Han,
  Matsui, Nagaosa, and Tokura}}]{bib-YU2010}
\bibinfo{author}{\bibfnamefont{X.~Z.} \bibnamefont{Yu}},
  \bibinfo{author}{\bibfnamefont{Y.}~\bibnamefont{Onose}},
  \bibinfo{author}{\bibfnamefont{N.~K. J. H.~P.} \bibnamefont{an~J.~H.~Han}},
  \bibinfo{author}{\bibfnamefont{Y.}~\bibnamefont{Matsui}},
  \bibinfo{author}{\bibfnamefont{N.}~\bibnamefont{Nagaosa}}, \bibnamefont{and}
  \bibinfo{author}{\bibfnamefont{Y.}~\bibnamefont{Tokura}},
  \bibinfo{journal}{\nature} \textbf{\bibinfo{volume}{465}},
  \bibinfo{pages}{901} (\bibinfo{year}{2010}).

\bibitem[{\citenamefont{Romming et~al.}(2013)\citenamefont{Romming, Hanneken,
  Menzel, Bickel, Wolter, von Bergmann, Kubetzka, and
  Wiesendanger}}]{bib-ROM2013}
\bibinfo{author}{\bibfnamefont{N.}~\bibnamefont{Romming}},
  \bibinfo{author}{\bibfnamefont{C.}~\bibnamefont{Hanneken}},
  \bibinfo{author}{\bibfnamefont{M.}~\bibnamefont{Menzel}},
  \bibinfo{author}{\bibfnamefont{J.~E.} \bibnamefont{Bickel}},
  \bibinfo{author}{\bibfnamefont{B.}~\bibnamefont{Wolter}},
  \bibinfo{author}{\bibfnamefont{K.}~\bibnamefont{von Bergmann}},
  \bibinfo{author}{\bibfnamefont{A.}~\bibnamefont{Kubetzka}}, \bibnamefont{and}
  \bibinfo{author}{\bibfnamefont{R.}~\bibnamefont{Wiesendanger}},
  \bibinfo{journal}{\science} \textbf{\bibinfo{volume}{341}},
  \bibinfo{pages}{636} (\bibinfo{year}{2013}).

\bibitem[{\citenamefont{Feldkeller}(1965)}]{bib-FEL1965}
\bibinfo{author}{\bibfnamefont{R.}~\bibnamefont{Feldkeller}},
  \bibinfo{journal}{\zap} \textbf{\bibinfo{volume}{19}}, \bibinfo{pages}{530}
  (\bibinfo{year}{1965}).

\bibitem[{\citenamefont{D\"{o}ring}(1968)}]{bib-DOE1968}
\bibinfo{author}{\bibfnamefont{W.}~\bibnamefont{D\"{o}ring}},
  \bibinfo{journal}{\jap} \textbf{\bibinfo{volume}{39}}, \bibinfo{pages}{1006}
  (\bibinfo{year}{1968}).

\bibitem[{\citenamefont{Malozemoff and Slonczewski}(1979)}]{bib-MAL1979}
\bibinfo{author}{\bibfnamefont{A.~P.} \bibnamefont{Malozemoff}}
  \bibnamefont{and} \bibinfo{author}{\bibfnamefont{J.~C.}
  \bibnamefont{Slonczewski}}, \emph{\bibinfo{title}{Magnetic domain walls in
  bubble materials}} (\bibinfo{publisher}{Academic press},
  \bibinfo{year}{1979}).

\bibitem[{\citenamefont{Thiaville et~al.}(2003)\citenamefont{Thiaville,
  {Garc\'{\i}a}, Dittrich, Miltat, and Schrefl}}]{bib-THI2003}
\bibinfo{author}{\bibfnamefont{A.}~\bibnamefont{Thiaville}},
  \bibinfo{author}{\bibfnamefont{J.~M.} \bibnamefont{{Garc\'{\i}a}}},
  \bibinfo{author}{\bibfnamefont{R.}~\bibnamefont{Dittrich}},
  \bibinfo{author}{\bibfnamefont{J.}~\bibnamefont{Miltat}}, \bibnamefont{and}
  \bibinfo{author}{\bibfnamefont{T.}~\bibnamefont{Schrefl}},
  \bibinfo{journal}{\prb} \textbf{\bibinfo{volume}{67}},
  \bibinfo{pages}{094410} (\bibinfo{year}{2003}).

\bibitem[{\citenamefont{Kim and Tchernyshyov}(2013)}]{bib-KIM2013}
\bibinfo{author}{\bibfnamefont{S.~K.} \bibnamefont{Kim}} \bibnamefont{and}
  \bibinfo{author}{\bibfnamefont{O.}~\bibnamefont{Tchernyshyov}},
  \bibinfo{journal}{\prb} \textbf{\bibinfo{volume}{88}},
  \bibinfo{pages}{174402} (\bibinfo{year}{2013}).

\bibitem[{\citenamefont{Piao et~al.}(2013)\citenamefont{Piao, Shim, Djuhana,
  and Kim}}]{bib-PIA2013}
\bibinfo{author}{\bibfnamefont{H.-G.} \bibnamefont{Piao}},
  \bibinfo{author}{\bibfnamefont{J.-H.} \bibnamefont{Shim}},
  \bibinfo{author}{\bibfnamefont{D.}~\bibnamefont{Djuhana}}, \bibnamefont{and}
  \bibinfo{author}{\bibfnamefont{D.-H.} \bibnamefont{Kim}},
  \bibinfo{journal}{\apl} \textbf{\bibinfo{volume}{102}},
  \bibinfo{pages}{112405} (\bibinfo{year}{2013}).

\bibitem[{\citenamefont{Forster et~al.}(2002)\citenamefont{Forster, Schrefl,
  Suess, Scholz, Tsiantos, Dittrich, and Fidler}}]{bib-FOR2002}
\bibinfo{author}{\bibfnamefont{H.}~\bibnamefont{Forster}},
  \bibinfo{author}{\bibfnamefont{T.}~\bibnamefont{Schrefl}},
  \bibinfo{author}{\bibfnamefont{D.}~\bibnamefont{Suess}},
  \bibinfo{author}{\bibfnamefont{W.}~\bibnamefont{Scholz}},
  \bibinfo{author}{\bibfnamefont{V.}~\bibnamefont{Tsiantos}},
  \bibinfo{author}{\bibfnamefont{R.}~\bibnamefont{Dittrich}}, \bibnamefont{and}
  \bibinfo{author}{\bibfnamefont{J.}~\bibnamefont{Fidler}},
  \bibinfo{journal}{\jap} \textbf{\bibinfo{volume}{91}}, \bibinfo{pages}{6914}
  (\bibinfo{year}{2002}).

\bibitem[{\citenamefont{Thiaville and Nakatani}(2006)}]{bib-THI2006}
\bibinfo{author}{\bibfnamefont{A.}~\bibnamefont{Thiaville}} \bibnamefont{and}
  \bibinfo{author}{\bibfnamefont{Y.}~\bibnamefont{Nakatani}},
  \emph{\bibinfo{title}{Spin dynamics in confined magnetic structures III}}
  (\bibinfo{publisher}{Springer}, \bibinfo{address}{Berlin},
  \bibinfo{year}{2006}), vol. \bibinfo{volume}{101} of
  \emph{\bibinfo{series}{Topics Appl. Physics}}, \bibinfo{type}{theory,
  simulation, magnetism, review} \bibinfo{chapter}{Domain-wall dynamics in
  nanowires and nanostrips}, pp. \bibinfo{pages}{161--206}.

\bibitem[{\citenamefont{Yan et~al.}(2011)\citenamefont{Yan, Andreas, Kakay,
  Garcia-Sanchez, and Hertel}}]{bib-YAN2011b}
\bibinfo{author}{\bibfnamefont{M.}~\bibnamefont{Yan}},
  \bibinfo{author}{\bibfnamefont{C.}~\bibnamefont{Andreas}},
  \bibinfo{author}{\bibfnamefont{A.}~\bibnamefont{Kakay}},
  \bibinfo{author}{\bibfnamefont{F.}~\bibnamefont{Garcia-Sanchez}},
  \bibnamefont{and} \bibinfo{author}{\bibfnamefont{R.}~\bibnamefont{Hertel}},
  \bibinfo{journal}{\apl} \textbf{\bibinfo{volume}{99}},
  \bibinfo{pages}{122505} (\bibinfo{year}{2011}).

\bibitem[{\citenamefont{Thiaville and Nakatani}(2009)}]{bib-THI2008}
\bibinfo{author}{\bibfnamefont{A.}~\bibnamefont{Thiaville}} \bibnamefont{and}
  \bibinfo{author}{\bibfnamefont{Y.}~\bibnamefont{Nakatani}},
  \emph{\bibinfo{title}{Nanomagnetism and Spintronics}}
  (\bibinfo{publisher}{Elsevier}, \bibinfo{year}{2009}), \bibinfo{type}{review,
  simulation, magnetism} \bibinfo{chapter}{Micromagnetic simulation of domain
  wall dynamics in nanostrips}.

\bibitem[{\citenamefont{Lee et~al.}(2006)\citenamefont{Lee, Ji, G\"{o}sele, and
  Nielsch}}]{bib-LEE2006}
\bibinfo{author}{\bibfnamefont{W.}~\bibnamefont{Lee}},
  \bibinfo{author}{\bibfnamefont{R.}~\bibnamefont{Ji}},
  \bibinfo{author}{\bibfnamefont{U.}~\bibnamefont{G\"{o}sele}}, \bibnamefont{and}
  \bibinfo{author}{\bibfnamefont{K.}~\bibnamefont{Nielsch}},
  \bibinfo{journal}{\NatMater} \textbf{\bibinfo{volume}{5}},
  \bibinfo{pages}{741} (\bibinfo{year}{2006}).

\bibitem[{\citenamefont{Lee et~al.}(2008)\citenamefont{Lee, Schwirn, Steinhart,
  Pippel, Scholz, and G\"{o}sele}}]{bib-LEE2008}
\bibinfo{author}{\bibfnamefont{W.}~\bibnamefont{Lee}},
  \bibinfo{author}{\bibfnamefont{K.}~\bibnamefont{Schwirn}},
  \bibinfo{author}{\bibfnamefont{M.}~\bibnamefont{Steinhart}},
  \bibinfo{author}{\bibfnamefont{E.}~\bibnamefont{Pippel}},
  \bibinfo{author}{\bibfnamefont{R.}~\bibnamefont{Scholz}}, \bibnamefont{and}
  \bibinfo{author}{\bibfnamefont{U.}~\bibnamefont{G\"{o}sele}},
  \bibinfo{journal}{\NatNanotech} \textbf{\bibinfo{volume}{3}},
  \bibinfo{pages}{234} (\bibinfo{year}{2008}).

\bibitem[{\citenamefont{Locatelli et~al.}(2006)\citenamefont{Locatelli, Aballe,
  Mente{\c{s}}, Kiskinova, and Bauer}}]{bib-LOC2006}
\bibinfo{author}{\bibfnamefont{A.}~\bibnamefont{Locatelli}},
  \bibinfo{author}{\bibfnamefont{L.}~\bibnamefont{Aballe}},
  \bibinfo{author}{\bibfnamefont{T.~O.} \bibnamefont{Mente{\c{s}}}},
  \bibinfo{author}{\bibfnamefont{M.}~\bibnamefont{Kiskinova}},
  \bibnamefont{and} \bibinfo{author}{\bibfnamefont{E.}~\bibnamefont{Bauer}},
  \bibinfo{journal}{\sia} \textbf{\bibinfo{volume}{38}}, \bibinfo{pages}{12}
  (\bibinfo{year}{2006}).

\bibitem[{\citenamefont{Alouges et~al.}(2012)\citenamefont{Alouges, Kritsikis,
  and Toussaint}}]{bib-ALO2012}
\bibinfo{author}{\bibfnamefont{F.}~\bibnamefont{Alouges}},
  \bibinfo{author}{\bibfnamefont{E.}~\bibnamefont{Kritsikis}},
  \bibnamefont{and} \bibinfo{author}{\bibfnamefont{J.-C.}
  \bibnamefont{Toussaint}}, \bibinfo{journal}{\PhysicaB}
  \textbf{\bibinfo{volume}{407}}, \bibinfo{pages}{1345} (\bibinfo{year}{2012}).

\bibitem[{\citenamefont{Fert and Piraux}(1999)}]{bib-FER1999a}
\bibinfo{author}{\bibfnamefont{A.}~\bibnamefont{Fert}} \bibnamefont{and}
  \bibinfo{author}{\bibfnamefont{J.~L.} \bibnamefont{Piraux}},
  \bibinfo{journal}{\jmmm} \textbf{\bibinfo{volume}{200}}, \bibinfo{pages}{338}
  (\bibinfo{year}{1999}).

\bibitem[{\citenamefont{Wang et~al.}(2009)\citenamefont{Wang, Wang, Fu,
  Hasegawa, Li, Saito, and Ishio}}]{bib-WAN2009}
\bibinfo{author}{\bibfnamefont{T.}~\bibnamefont{Wang}},
  \bibinfo{author}{\bibfnamefont{Y.}~\bibnamefont{Wang}},
  \bibinfo{author}{\bibfnamefont{Y.}~\bibnamefont{Fu}},
  \bibinfo{author}{\bibfnamefont{T.}~\bibnamefont{Hasegawa}},
  \bibinfo{author}{\bibfnamefont{F.~S.} \bibnamefont{Li}},
  \bibinfo{author}{\bibfnamefont{H.}~\bibnamefont{Saito}}, \bibnamefont{and}
  \bibinfo{author}{\bibfnamefont{S.}~\bibnamefont{Ishio}},
  \bibinfo{journal}{\Nanotech} \textbf{\bibinfo{volume}{20}},
  \bibinfo{pages}{105707} (\bibinfo{year}{2009}).

\bibitem[{\citenamefont{Pitzschel et~al.}(2011)\citenamefont{Pitzschel,
  Bachmann, Martens, Montero-Moreno, Kimling, Meier, Escrig, Nielsch, and
  G\"{o}rlitz}}]{bib-PIT2011}
\bibinfo{author}{\bibfnamefont{K.}~\bibnamefont{Pitzschel}},
  \bibinfo{author}{\bibfnamefont{J.}~\bibnamefont{Bachmann}},
  \bibinfo{author}{\bibfnamefont{S.}~\bibnamefont{Martens}},
  \bibinfo{author}{\bibfnamefont{J.~M.} \bibnamefont{Montero-Moreno}},
  \bibinfo{author}{\bibfnamefont{J.}~\bibnamefont{Kimling}},
  \bibinfo{author}{\bibfnamefont{G.}~\bibnamefont{Meier}},
  \bibinfo{author}{\bibfnamefont{J.}~\bibnamefont{Escrig}},
  \bibinfo{author}{\bibfnamefont{K.}~\bibnamefont{Nielsch}}, \bibnamefont{and}
  \bibinfo{author}{\bibfnamefont{D.}~\bibnamefont{G\"{o}rlitz}},
  \bibinfo{journal}{\jap} \textbf{\bibinfo{volume}{109}},
  \bibinfo{pages}{033907} (\bibinfo{year}{2011}).

\bibitem[{\citenamefont{Kimling et~al.}(2011)\citenamefont{Kimling, Kronast,
  Martens, B\"ohnert, Martens, Herrero-Albillos, Tati-Bismaths, Merkt, Nielsch,
  and Meier}}]{bib-KIM2011b}
\bibinfo{author}{\bibfnamefont{J.}~\bibnamefont{Kimling}},
  \bibinfo{author}{\bibfnamefont{F.}~\bibnamefont{Kronast}},
  \bibinfo{author}{\bibfnamefont{S.}~\bibnamefont{Martens}},
  \bibinfo{author}{\bibfnamefont{T.}~\bibnamefont{B\"ohnert}},
  \bibinfo{author}{\bibfnamefont{M.}~\bibnamefont{Martens}},
  \bibinfo{author}{\bibfnamefont{J.}~\bibnamefont{Herrero-Albillos}},
  \bibinfo{author}{\bibfnamefont{L.}~\bibnamefont{Tati-Bismaths}},
  \bibinfo{author}{\bibfnamefont{U.}~\bibnamefont{Merkt}},
  \bibinfo{author}{\bibfnamefont{K.}~\bibnamefont{Nielsch}}, \bibnamefont{and}
  \bibinfo{author}{\bibfnamefont{G.}~\bibnamefont{Meier}},
  \bibinfo{journal}{\prb} \textbf{\bibinfo{volume}{84}},
  \bibinfo{pages}{174406} (\bibinfo{year}{2011}).

\bibitem[{\citenamefont{Biziere et~al.}(2013)\citenamefont{Biziere, Gatel,
  Lassalle-Balier, Clochard, Wegrowe, and Snoeck}}]{bib-BIZ2013}
\bibinfo{author}{\bibfnamefont{N.}~\bibnamefont{Biziere}},
  \bibinfo{author}{\bibfnamefont{C.}~\bibnamefont{Gatel}},
  \bibinfo{author}{\bibfnamefont{R.}~\bibnamefont{Lassalle-Balier}},
  \bibinfo{author}{\bibfnamefont{M.~C.} \bibnamefont{Clochard}},
  \bibinfo{author}{\bibfnamefont{J.~E.} \bibnamefont{Wegrowe}},
  \bibnamefont{and} \bibinfo{author}{\bibfnamefont{E.}~\bibnamefont{Snoeck}},
  \bibinfo{journal}{\nanolett} \textbf{\bibinfo{volume}{13}},
  \bibinfo{pages}{2053} (\bibinfo{year}{2013}).

\bibitem[{\citenamefont{Moore et~al.}(2008)\citenamefont{Moore, Miron, Gaudin,
  Serret, Rodmacq, Schuhl, Pizzini, Vogel, and Bonfim}}]{bib-MOO2008}
\bibinfo{author}{\bibfnamefont{T.~A.} \bibnamefont{Moore}},
  \bibinfo{author}{\bibfnamefont{M.}~\bibnamefont{Miron}},
  \bibinfo{author}{\bibfnamefont{G.}~\bibnamefont{Gaudin}},
  \bibinfo{author}{\bibfnamefont{G.}~\bibnamefont{Serret}},
  \bibinfo{author}{\bibfnamefont{S.~A.~B.} \bibnamefont{Rodmacq}},
  \bibinfo{author}{\bibfnamefont{A.}~\bibnamefont{Schuhl}},
  \bibinfo{author}{\bibfnamefont{S.}~\bibnamefont{Pizzini}},
  \bibinfo{author}{\bibfnamefont{J.}~\bibnamefont{Vogel}}, \bibnamefont{and}
  \bibinfo{author}{\bibfnamefont{M.}~\bibnamefont{Bonfim}},
  \bibinfo{journal}{\apl} \textbf{\bibinfo{volume}{93}},
  \bibinfo{pages}{262504} (\bibinfo{year}{2008}).

\bibitem[{\citenamefont{Miron et~al.}(2011)\citenamefont{Miron, Moore,
  Szambolics, Buda-Prejbeanu, Auffret, Rodmacq, Pizzini, Vogel, Bonfim, Schuhl,
  and Gaudin}}]{bib-MIR2011}
\bibinfo{author}{\bibfnamefont{I.~M.} \bibnamefont{Miron}},
  \bibinfo{author}{\bibfnamefont{T.}~\bibnamefont{Moore}},
  \bibinfo{author}{\bibfnamefont{H.}~\bibnamefont{Szambolics}},
  \bibinfo{author}{\bibfnamefont{L.~D.} \bibnamefont{Buda-Prejbeanu}},
  \bibinfo{author}{\bibfnamefont{S.}~\bibnamefont{Auffret}},
  \bibinfo{author}{\bibfnamefont{B.}~\bibnamefont{Rodmacq}},
  \bibinfo{author}{\bibfnamefont{S.}~\bibnamefont{Pizzini}},
  \bibinfo{author}{\bibfnamefont{J.}~\bibnamefont{Vogel}},
  \bibinfo{author}{\bibfnamefont{M.}~\bibnamefont{Bonfim}},
  \bibinfo{author}{\bibfnamefont{A.}~\bibnamefont{Schuhl}}, \bibnamefont{and}
  \bibinfo{author}{\bibfnamefont{G.}~\bibnamefont{Gaudin}},
  \bibinfo{journal}{\NatMater} \textbf{\bibinfo{volume}{10}},
  \bibinfo{pages}{419} (\bibinfo{year}{2011}).

\bibitem[{\citenamefont{Yan et~al.}(2010)\citenamefont{Yan, K\'akay, Gliga, and
  Hertel}}]{bib-YAN2010}
\bibinfo{author}{\bibfnamefont{M.}~\bibnamefont{Yan}},
  \bibinfo{author}{\bibfnamefont{A.}~\bibnamefont{K\'akay}},
  \bibinfo{author}{\bibfnamefont{S.}~\bibnamefont{Gliga}}, \bibnamefont{and}
  \bibinfo{author}{\bibfnamefont{R.}~\bibnamefont{Hertel}},
  \bibinfo{journal}{\prl} \textbf{\bibinfo{volume}{104}},
  \bibinfo{pages}{057201} (\bibinfo{year}{2010}).

\bibitem[{\citenamefont{Parkin}()}]{bib-PAR2004}
\bibinfo{author}{\bibfnamefont{S.~S.~P.} \bibnamefont{Parkin}},
  \emph{\bibinfo{title}{U.s. patents 6834005, 6898132, 6920062}}.

\bibitem[{\citenamefont{Parkin et~al.}(2008)\citenamefont{Parkin, Hayashi, and
  Thomas}}]{bib-PAR2008}
\bibinfo{author}{\bibfnamefont{S.~S.~P.} \bibnamefont{Parkin}},
  \bibinfo{author}{\bibfnamefont{M.}~\bibnamefont{Hayashi}}, \bibnamefont{and}
  \bibinfo{author}{\bibfnamefont{L.}~\bibnamefont{Thomas}},
  \bibinfo{journal}{\science} \textbf{\bibinfo{volume}{320}},
  \bibinfo{pages}{190} (\bibinfo{year}{2008}).

\end{thebibliography}

\end{document}